\documentclass{PoS}
\usepackage{amsmath,amssymb,bm,float,bbold,multirow,graphicx}
\usepackage{cite}

\usepackage{caption}
\DeclareCaptionLabelFormat{andtable}{#1~#2  \&  \tablename~\thetable}

\usepackage{xcolor,colortbl}
\definecolor{Gray}{gray}{0.95}
\definecolor{RGray}{gray}{0.85}
\definecolor{CGray}{gray}{0.92}

\title{A class of Z$\mathbf{^{\,\prime}}$ models with non-universal couplings and protected flavor-changing interactions}

\ShortTitle{A class of $Z^{\,\prime}$ models with non-universal couplings}

\author{\speaker{Javier Fuentes-Mart\'in}\\%
        IFIC, Universitat de Val\`encia-CSIC\\
        Apt. Correus 22085, E-46071 Val\`encia, Spain\\
        E-mail: \email{javier.fuentes@ific.uv.es}}

\abstract{
Motivated by the $b\to s\ell^+\ell^-$ anomalies recently reported by the LHCb collaboration, I present a class of flavored $\mathrm{U(1)}^\prime$ gauge extensions of the Standard Model that naturally accommodates them and possesses a rich phenomenology. This model is characterized by the presence of tree-level flavor-changing interactions in the down-quark sector, protected by off-diagonal quark-mixing matrix elements. Anomaly cancellation fixes the extension of the symmetry to the lepton sector in a very specific way, giving rise to flavor-conserving family-non-universal $Z^{\,\prime}$ couplings. The fermion content of this model is the same as in the SM while the scalar sector is extended with an extra Higgs doublet and a scalar singlet. The model will be tested in the next run of LHC and presents specific correlations in certain flavor observables that allow to clearly discriminate among them and from other new physics signals. 
}

\FullConference{The European Physical Society Conference on High Energy Physics\\
		 22-29 July 2015\\
		 Vienna, Austria}

\begin{document}

\section{Introduction}\label{Sec:intro}
The great success of the LHC during its first run has provided a plethora of data that have tested the Standard Model (SM) to great accuracy. The high precision achieved in many observables, together with their agreement with the SM predictions, has resulted in strong implications for new physics (NP) frameworks, increasing the NP scale or requiring non-trivial flavor structures. In spite of the undisputed success of the SM predictions, this run of LHC has left several hints of NP in semileptonic transitions $b\to s\ell^+\ell^-$. In particular the recent measurement by the LHCb collaboration of the ratio $R_K = \mathrm{Br}(B \rightarrow K \mu^+ \mu^-)/\mathrm{Br}(B\rightarrow K e^+ e^-)$ shows a deviation from the SM prediction at the $2.6\sigma$ level, hinting to a large violation of lepton flavor universality~\cite{Aaij:2014ora}. Several global analyses of $b\to s\ell^+\ell^-$ transitions have been performed~\cite{Descotes-Genon:2013wba,Altmannshofer:2014rta,Altmannshofer:2015sma,Descotes-Genon:2015uva}, showing a significant preference towards a NP explanation of the experimental anomalies found in these transitions. Among the many observables entering in the fit, the angular analysis of $B\to K^*\mu^+\mu^-$ decays, also by the LHCb collaboration, presents a clear example of deviation from the SM prediction in the observable $P'_5$ with $2.9\sigma$ significance in two of the bins~\cite{Aaij:2013qta}.

The NP necessary to accommodate the $b\to s\ell^+\ell^-$ anomalies should be non-universal in the lepton sector and present flavor changing neutral currents (FCNCs) in the down-quark sector. Various attempts to analize these anomalies in a model-independent way or to accommodate them with specific NP models can be found the literature~\cite{Altmannshofer:2013foa,Buras:2013dea,Gauld:2013qja,Altmannshofer:2014cfa,Crivellin:2015mga,
Crivellin:2015lwa,Sierra:2015fma,Buras:2013qja,Gauld:2013qba,Crivellin:2015era,Alonso:2014csa,Hiller:2014yaa,Varzielas:2015iva,Gripaios:2014tna,Jager:2014rwa,Niehoff:2015bfa,Becirevic:2015asa,Falkowski:2015zwa,Alonso:2015sja,Belanger:2015nma}. In this talk I will present a $\mathrm{U(1)}^\prime$ gauge symmetry implementation with a minimal particle content and characterized by having all the flavor violations controlled by the gauge symmetry, which makes them proportional to off-diagonal elements of the Cabibbo-Kobayashi-Maskawa (CKM) matrix. This model can be obtained by gauging the global symmetry introduced by Branco, Grimus and Lavoura (BGL) in the context of two-Higgs-doublet models (2HDMs) to address the flavor problem of these models while allowing for (controlled) flavor violations~\cite{Branco:1996bq}, providing a solution completely different from the hypothesis of natural flavor conservation~\cite{Glashow:1976nt,Paschos:1976ay}. 

The outline of the talk is as follows: In Section~\ref{Sec:BGL} I introduce the BGL models and their main properties. Section~\ref{Sec:model} is devoted to the construction of the gauged $\mathrm{U(1)}_{\mbox{\scriptsize BGL}}$ model. A discussion on the main constraints and phenomenological implications of the new gauge sector is given in Section~\ref{Sec:pheno}. I summarize in Section~\ref{Sec:conclusions}.

\section{The Branco-Grimus-Lavoura model}\label{Sec:BGL}
BGL models \cite{Branco:1996bq} provide a class of 2HDMs characterized by the presence of FCNCs at tree-level entirely fixed by CKM matrix elements and the ratio of vevs of the Higgs doublets. This is achieved by the imposition of a global horizontal symmetry that gives rise to a specific set of Yukawa textures. The Yukawa sector of the model is given by
\begin{align}\label{eq:Lagquarks}
-\mathcal{L}^{\mbox{\scriptsize{quark}}}_{\mbox{\scriptsize{Yuk}}}&=\overline{q_L^0}\, \Gamma_i\,
\Phi_i d^0_R+\overline{q_L^0}\, \Delta_i\, \widetilde{\Phi}_i  u^0_R+\text{h.c.} \,,
\end{align}  
where $\tilde \Phi_i\equiv i\sigma_2\Phi_i^*$, with $\sigma_2$ the Pauli matrix. Both Higgs doublets, $\Phi_i$ ($i=1,2$), acquire vacuum expectation values (vev) $|\langle\Phi^0_i\rangle|=v_i/\sqrt{2}$ with $v\equiv \left(v_1^2+v_2^2\right) = (\sqrt{2}  G_F)^{-1/2}\simeq 246$~GeV fixed by measurements of the muon lifetime; as usual I define $\tan \beta = v_2/v_1$. In this talk I will focus on the so-called top-BGL implementation where up and down Yukawa matrices, $\Delta_i$ and $\Gamma_i$, are constrained by the BGL symmetry to have the following structure:
%%%%%
\begin{align}\label{eq:BGLText}
\begin{aligned}
\Gamma_1&=
\begin{pmatrix}
\times & \times & \times\\
\times & \times & \times\\
0 & 0 & 0
\end{pmatrix}\,,\quad
\Gamma_2=
\begin{pmatrix}
0 & 0 & 0\\
0 & 0 & 0\\
\times & \times & \times
\end{pmatrix}\,, \\[0.2cm]
\Delta_1&=
\begin{pmatrix}
\times & \times & 0\\
\times & \times & 0\\
0 & 0 & 0
\end{pmatrix} \,\,,\quad
\Delta_2=
\begin{pmatrix}
0 & 0 & 0\\
0 & 0 & 0\\
0 & 0 & \times
\end{pmatrix} \,. 
\end{aligned}
\end{align}
%%%%%
These quark textures introduce FCNCs only in the down-quark sector that are suppressed by quark masses and off-diagonal elements of the third row of the CKM matrix~\cite{Branco:1996bq}, which results in a strong suppression of flavor-changing processes involving light quarks. This symmetry suppression of FCNCs allows top-BGL models to avoid experimental constraints even when the new scalars remain light, with masses of $\mathcal{O}\left(100\right)$~GeV~\cite{Botella:2014ska,Bhattacharyya:2014nja}.

Given an Abelian symmetry characterized by the field transformation
\begin{align}
\psi\rightarrow e^{iQ^{\psi}}\psi\,,
\end{align}
the most general implementation of the top-BGL Yukawa textures is defined by the following set of charges
\begin{align}
\begin{aligned}  \label{eq:system}
Q_L^q&=\frac{1}{2}\left[\mathrm{diag}\left(X_{uR},X_{uR},X_{tR}\right)+X_{dR}\,\mathbb{1}\right]\,,\\
Q_R^u&=\mathrm{diag}\left(X_{uR},X_{uR},X_{tR}\right)\,,\\
Q_R^d&=X_{dR}\,\mathbb{1}\,,\\
Q^\Phi&= \mathrm{diag}\,(X_{\Phi_1}, X_{\Phi_2})= \frac{1}{2}\mathrm{diag}\,\left(X_{uR}-X_{dR},X_{tR}-X_{dR}\right)\,,
\end{aligned}
\end{align}
with $X_{uR}\neq X_{tR}$.

Although the Abelian BGL symmetry can be discrete, it always leads to an enhanced accidental $\mathrm{U(1)}$ global symmetry in the scalar sector, which results in the presence of an undesired Goldstone in the theory~\cite{Branco:1996bq}. Several solutions to this problem of BGL models can be found in Refs.~\cite{Branco:1996bq,Celis:2014iua,Celis:2014jua}. Following Ref.~\cite{Celis:2015ara} I present here a different solution based on the promotion of the BGL symmetry to a local one. This way the gauging of the BGL symmetry serves a two-fold purpose: it provides a natural solution to the Goldstone boson problem in BGL models and introduces at the same time a new gauge boson with a very rich phenomenology, allowing for an explanation of the $b\to s\ell^+\ell^-$ anomalies in terms of symmetry principles.

\section{Gauged BGL symmetry}\label{Sec:model}
In this section I extend the SM gauge symmetry with an extra $\mathrm{U(1)}^\prime$ factor that is identified with the BGL symmetry introduced in the previous section. The scalar sector of this model consists of two Higgs doublets and a complex SM singlet, necessary to give a heavy mass to the new gauge boson, while the fermion content remains the same as in the SM. As we are dealing with a chiral symmetry, one should pay special attention to the cancellation of anomalies when gauging the BGL symmetry. In Ref.~\cite{Celis:2014iua} it was shown that in BGL 2HDMs the cancellation is automatic for the QCD currents, i.e. $\mathrm{U(1)}^\prime[\mathrm{SU(3)}_C]^2$. However, this is not the case for the rest of the anomalies,
\begin{align}\label{eq:anomalies}
\begin{aligned}
&\mathrm{U(1)}^\prime [\mathrm{SU(2)}_L]^2\,,\quad \mathrm{U(1)}^\prime [\mathrm{U(1)}_Y]^2\,, 
\qquad [\mathrm{U(1)}^\prime]^2 \mathrm{U(1)}_Y \,,\\
&[\mathrm{U(1)}^\prime]^3\,,\quad  \qquad \quad \,  \mathrm{U(1)}^\prime [\mbox{Gravity}]^2 \,.
\end{aligned}
\end{align}
In particular we find that, with the charge assignments in Eq.~\eqref{eq:system}, there is no solution to all the anomaly cancellation conditions unless we extend the symmetry to the lepton sector. Just like in the SM we find that anomaly cancellation can only be granted through the interplay of quarks and leptons. Taking the most general symmetry implementation, with all the lepton charges being free parameters,
\begin{align}
Q_L^\ell&=\mathrm{diag}\left(X_{eL},X_{\mu L},X_{\tau L}\right)\,,\qquad Q_R^e=\mathrm{diag}\left(X_{eR},X_{\mu R},X_{\tau R}\right)\,,
\end{align}
we find only one solution to the anomaly cancellation conditions. It is characterized by only two free charges, $X_{dR}$ and $X_{\mu R}$, up to lepton-flavor permutations:
\begin{align}
\begin{split}
X_{uR}&=-X_{dR}-\frac{1}{3}X_{\mu R}\,,\quad X_{tR}=-4X_{dR}+\frac{2}{3}X_{\mu R}\,,\\
X_{e L}&=X_{dR}+\frac{1}{6}X_{\mu R}\,,\quad\hspace{6.5pt} X_{e R}=2X_{dR}+\frac{1}{3}X_{\mu R}\,,\\
X_{\tau L}&=\frac{9}{2}X_{dR}-X_{\mu R}\,,\quad\hspace{6pt} X_{\tau R}= 7 X_{dR}-\frac{4}{3}X_{\mu R}\,,\\
X_{\mu_L}&=-X_{dR}+\frac{5}{6}X_{\mu R}\,.
\end{split}
\end{align}
However, one should note that the global scale of the charges is unphysical and only accounts for a rescaling of the $\mathrm{U(1)}_{\mbox{\scriptsize BGL}}$ gauge coupling, $g^\prime$, allowing us to freely remove one of the charges. As we can see, anomaly cancellation conditions determine the extension of the symmetry to the lepton sector in an unique way, with the charged-lepton Yukawa sector of the model taking the form
\begin{align}
-\mathcal{L}^{\mbox{\scriptsize c-leptons}}_{\mbox{\scriptsize{Yuk}}}&=\overline{\ell_L^0}\, \Pi_i\, \Phi_i e^0_R+\text{h.c.} \,,
\end{align}
where
\begin{equation}
\Pi_1=
\begin{pmatrix}
\times&0&0\\
0&\times&0\\
0&0&0
\end{pmatrix}\,,\quad \Pi_2=
\begin{pmatrix}
0&0&0\\
0&0&0\\
0&0&\times
\end{pmatrix}\,.
\end{equation}
Since the only source of flavor violation of the model is found in the Yukawa matrices, charged-lepton flavor conservation appears in this model as a natural consequence of the gauge symmetry. I call to attention that the neutrino sector of the model has not been specified. Extensions to account for neutrino masses and mixings can potentially modify the anomaly conditions, opening the possibility for new solutions. A systematic study of the neutrino sector will be presented in a future publication.

For phenomenological purposes it is convenient to eliminate the remaining freedom in the model by fixing $X_{\Phi_2}=0$ (or equivalently $X_{dR}=2/15\, X_{\mu R}$), so that the mixing between neutral gauge bosons is suppressed for large $\tan \beta$. For simplicity, I work in this limit and neglect mixing effects for the rest of the talk, leaving a more general analysis of the model for future work. Finally, without loss of generality, I choose a charge normalization by setting $X_{dR}=1$, with no physical implications. With these choices the $\mathrm{U(1)}_{\mbox{\scriptsize BGL}}$ charges read
\begin{align}  \label{eq:charges_model}
\begin{aligned}
Q_R^d&=\mathbb{1}\,,\quad
&Q_R^u&=\text{diag}\left(-\frac{7}{2},-\frac{7}{2},1\right)\,,\\
Q_L^q&=\text{diag}\left(-\frac{5}{4},-\frac{5}{4},1\right)\,,\quad
&Q_L^\ell&=\text{diag}\left(\frac{9}{4},\frac{21}{4},-3\right)\,,\\
Q_R^{e}&=\text{diag}\left(\frac{9}{2},\frac{15}{2},-3\right)\,,\quad
&Q^{\Phi}&=\text{diag}\left(-\frac{9}{4},0\right)\,.
\end{aligned}
\end{align}
Permutations of lepton flavors yield six different implementations of the symmetry which we denote as $(e,\mu,\tau)=(i,j,k)$, with the model implementation in Eq.~\eqref{eq:charges_model} denoted as $(1,2,3)$.

To avoid experimental constraints the new gauge boson associated to the BGL symmetry, $Z^{\,\prime}$, should have a heavy mass of a few TeV. This is achieved through the inclusion of a complex scalar SM singlet, $S$, charged under the new symmetry, that acquires a large vev $|\langle S \rangle | = v_S/\sqrt{2}\gg v$ and spontaneously breaks the extra gauge symmetry. Also note that the charge of the singlet, $X_S$, should be fixed in terms of the other scalar charges in order to avoid undesired Goldstone bosons (for more details see Ref.~\cite{Celis:2015ara}), I choose $X_S=1/2\left(X_{\Phi_1}-X_{\Phi_2}\right)=-9/8$. The Lagrangian for the new gauge sector then reads
 \begin{align}
 \mathcal{L}_{Z^\prime}\simeq -\frac{1}{4}Z_{\mu\nu}^{\,\prime}Z^{\,\prime\,\mu\nu}+\left|D_\mu\,S\right|^2-V\left(S\right)-J^\mu_{Z^\prime}\,Z^{\,\prime}_\mu \,.
 \end{align}
Here $Z_{\mu\nu}^{\,\prime}$ is the $Z^{\,\prime}$ field-strength tensor and the $Z^{\,\prime}$ current is denoted as $J^\mu_{Z^\prime}$. Its fermionic piece takes the form
\begin{align}
J_{Z^\prime}^{\,\mu}\supset g^\prime\,\overline{\psi_{i}}\,\gamma^\mu \left[\widetilde Q^\psi_{L,ij}\,P_L+\widetilde Q^\psi_{R,ij}\,P_R\right] \psi_{j}\,,
\end{align}
where $g^\prime$ is the $Z^{\,\prime}$ gauge coupling and $\widetilde Q^\psi$ stands for the $Z^{\,\prime}$ charges (see Eq.~\eqref{eq:charges_model}) rotated to the fermion physical eigenbasis 
\begin{align}\label{eq:flavorstr}
\widetilde Q^\psi_{R} &= Q_{R}^{\psi}\,,\quad  
\widetilde Q^u_{L} =  Q_{L}^{q} \,, \quad \widetilde Q^\ell_{L} = 
Q_{L}^{\ell}
\,,\quad
\widetilde Q^d_{L} =  - \frac{5}{4} \mathbb{1} +\frac{9}{4}   \, \begin{pmatrix}
 |V_{td}|^2  &    V_{ts} V_{td}^*   &    V_{tb}  V_{td}^*      \\
  V_{td}  V_{ts}^*     &    |V_{ts}|^2   &   V_{tb} V_{ts}^*   \\
   V_{td}  V_{tb}^*  &     V_{ts}  V_{tb}^*&     |V_{tb}|^2 
\end{pmatrix}       \,.
\end{align}
Note that $Z^{\,\prime}$-mediated flavor violations are only present in the left-handed down-quark sector.

\section{Phenomenological constraints and model predictions}\label{Sec:pheno}
In this section I will only highlight the main constraints and predictions concerning $Z^{\,\prime}$ observables and refer the reader to Ref.~\cite{Celis:2015ara} for an extended discussion on the phenomenology of the model. Since all BGL charges are fixed, $Z^{\,\prime}$ observables are completely determined in terms of just two free parameters, the $Z^{\,\prime}$ mass and gauge coupling.

Constraints on the $Z^{\,\prime}$ from low energy observables are only sensitive to the combination of parameters $M_{Z^\prime}/g^\prime$. Bounds from $B_s$-mixing give the limit $M_{Z^\prime}/g^\prime\gtrsim16$~TeV at $95\%$ CL and we find that other low-energy constraints such as those from neutrino trident production, atomic parity violation, electric dipole moments or anomalous magnetic moments are always weaker than the one from $B_s$-mixing. Also interesting are the LHC bounds on direct searches for a $Z^{\,\prime}$ decaying into a pair of leptons, since they allow to disentangle the two free parameters. Using the model independent analysis provided by the CMS collaboration~\cite{Chatrchyan:2012it,Khachatryan:2014fba} we find the exclusion limit in the mass of the $Z^{\,\prime}$, $M_{Z^\prime}\gtrsim3-4$ TeV, depending on the model implementation. Additionally, requiring the gauge couplings to remain perturbative we obtain an upper limit on the the value of $g^\prime$. The model develops a Landau pole at the see-saw scale, $\Lambda_{\mbox{\scriptsize{LP}}}\gtrsim 10^{14}$~GeV, for $g^\prime\lesssim0.14$ while if we push the Landau pole to the Planck scale, $\Lambda_{\mbox{\scriptsize{LP}}}\gtrsim 10^{19}$~GeV, we find the limit $g^\prime\lesssim0.12$.

%%%%%%%%%%%%%%%%%%%%%%%%%%%%%%%%%%%%%%%%%%%%%%%%%%%%%%%%%%%%%%%%%%%%%%%%%%
\begin{table}
\begin{center}
\tabcolsep 0.03in
\begin{tabular}{|c|c|c|c|c|c|c|} \rowcolor{RGray}
\hline 
Model   & $C_{10}^{\mbox{\scriptsize{NP}} \mu}/C_{9}^{\mbox{\scriptsize{NP}}\mu}$ &  $C_{9}^{\mbox{\scriptsize{NP}}e}/C_{9}^{\mbox{\scriptsize{NP}}\mu}$  & $C_{10}^{\mbox{\scriptsize{NP}}e}/C_9^{\mbox{\scriptsize{NP}}\mu}$   &  $C_{9}^{\mbox{\scriptsize{NP}} \tau}/C_{9}^{\mbox{\scriptsize{NP}}\mu}$  & $C_{10}^{\mbox{\scriptsize{NP}}\tau}/C_9^{\mbox{\scriptsize{NP}}\mu}$ &    $\kappa_{9}^{\mu}$    \\[0.1cm] 
\hline
(1,2,3) & $3/17$  & $9/17$                 & $3/17$   & $-8/17$                 & $0$ & $-1.235$ \\[0.1cm] \rowcolor{CGray}  
(1,3,2) & $0$  & $-9/8$ & $-3/8$   & $-17/8$ & $-3/8$    &  $0.581$  \\[0.1cm] 
(2,1,3) & $1/3$  & $17/9$                 & $1/3$  & $-8/9$                 & $0$  & $-0.654$ \\[0.1cm] \rowcolor{CGray}
(2,3,1) & $0$ & $-17/8$ & $-3/8$    & $-9/8$ & $-3/8$    &  $0.581$  \\[0.1cm]   
(3,1,2) & $1/3$  & $-8/9$                 & $0$  & $17/9$                 & $1/3$ & $-0.654$ \\[0.1cm] \rowcolor{CGray}
(3,2,1) & $3/17$ & $-8/17$                 & $0$ & $9/17$                 & $3/17$     & $-1.235$ \\[0.1cm] 
\hline
\end{tabular}
\caption{\it \small Correlations among the NP contributions to the effective operators $\mathcal{O}_{9,10}^{\ell}$.}
\label{tab:correlations}
\end{center}
\end{table}
%%%%%%%%%%%%%%%%%%%%%%%%%%%%%%%%%%%%%%%%%%%%%%%%%%%%%%%%%%%%%%%%%%%%%%%%%%

I now turn to the $b\to s\ell^+\ell^-$ anomalies, the effective Hamiltonian for these transitions reads
\begin{align}
\mathcal{H}_{\mbox{\scriptsize eff}}=   - \frac{G_F}{\sqrt{2}}   \frac{\alpha}{\pi} V_{tb}V_{ts}^*  
\sum_{i}   \left( C_i^{\ell} \mathcal{O}^{\ell}_i  +  C_i^{\prime \ell} \mathcal{O}^{\prime \ell}_i 
\right),
\end{align}
with
\begin{align}
\begin{aligned}
\mathcal{O}^{\ell}_9&=\left(\overline{s}\gamma_\mu P_Lb\right)\left(\overline{\ell}\gamma^\mu
 \ell\right)\,,    &\mathcal{O}_9^{\prime \ell} &=  \left(\overline{s}\gamma_\mu P_Rb\right)\left(\overline{\ell}\gamma^\mu \ell\right)\,,\\[5pt] \mathcal{O}^{\ell}_{10}&= \left(\overline{s}\gamma_\mu P_Lb\right)\left(\overline{\ell}\gamma^\mu\gamma_5 \ell\right)\,,   &\mathcal{O}_{10}^{\prime \ell}&= \left(\overline{s}\gamma_\mu P_Rb\right)\left(\overline{\ell}\gamma^\mu\gamma_5 \ell\right)\,.
\end{aligned}
\end{align}
The SM contribution to these operators is $C_{9}^{\mbox{\scriptsize{SM}}} \simeq - C_{10}^{\mbox{\scriptsize{SM}}} \simeq  4.2\,\forall\,\ell$, with negligible contributions to the primed operators. Since right-handed quark currents are flavor conserving in our model, $\mathcal{O}_{9,10}^{\prime \ell}$ also receive negligible contributions from the $Z^{\,\prime}$. Its contribution to $\mathcal{O}_{9,10}^\ell$ is given by
\begin{align}
\begin{aligned}
C_{9}^{\mbox{\scriptsize{NP}}  \ell}&\simeq     - \frac{\pi }{\alpha V_{ts}^* V_{tb} }\; {\widetilde Q}^{\,d}_{L,sb}  \left({\widetilde Q}^{\,e}_{L,\ell \ell}+{\widetilde Q}^{\,e}_{R,\ell \ell}\right) \left(\frac{g^\prime v   }{M_{Z^\prime} }\right)^2  \,, \\  
C_{10}^{\mbox{\scriptsize{NP}}  \ell}&\simeq    \frac{\pi }{\alpha V_{ts}^* V_{tb} }\; {\widetilde Q}^{\,d}_{L,sb}  \left({\widetilde Q}^{\,e}_{L,\ell \ell}-{\widetilde Q}^{\,e}_{R,\ell \ell}\right) \left(\frac{g^\prime v   }{M_{Z^\prime} }\right)^2  \,,
\end{aligned}
\end{align}
where $C_{i}^{\ell} \equiv C_{i}^{\mbox{\scriptsize{SM}}}  + C_{i}^{\mbox{\scriptsize{NP}} \ell}$. The correlations among the different contributions is shown in Table~\ref{tab:correlations} where I also provide the value of $C_{9}^{\mbox{\scriptsize{NP}}\mu}$ as a function of $g^\prime/M_{Z^\prime}$, which is given in terms of the following normalization
\begin{align}
 C_9^{\mbox{\scriptsize{NP}} \mu}  \equiv  \kappa_9^{\mu}   \times 10^{4} \left(   \frac{ g^{\prime}
 v }{  M_{Z^{\prime}}  }  \right)^2=\kappa_9^{\mu}   \times605~{\rm TeV}^2 \left(   \frac{ g^{\prime}
 }{  M_{Z^{\prime}}  }  \right)^2 \,.
\end{align}
%

%%%%%%%%%%%%%%%%%%%%%%%%%%%%%%%%%%%%%%%%%%%%%%%%%%%%%%%%%%%%%%%%%%%%%%%%%%
\begin{figure}[ht]
\begin{minipage}{0.49\textwidth}
\centering
\includegraphics[width=7.5cm,height=7.5cm]{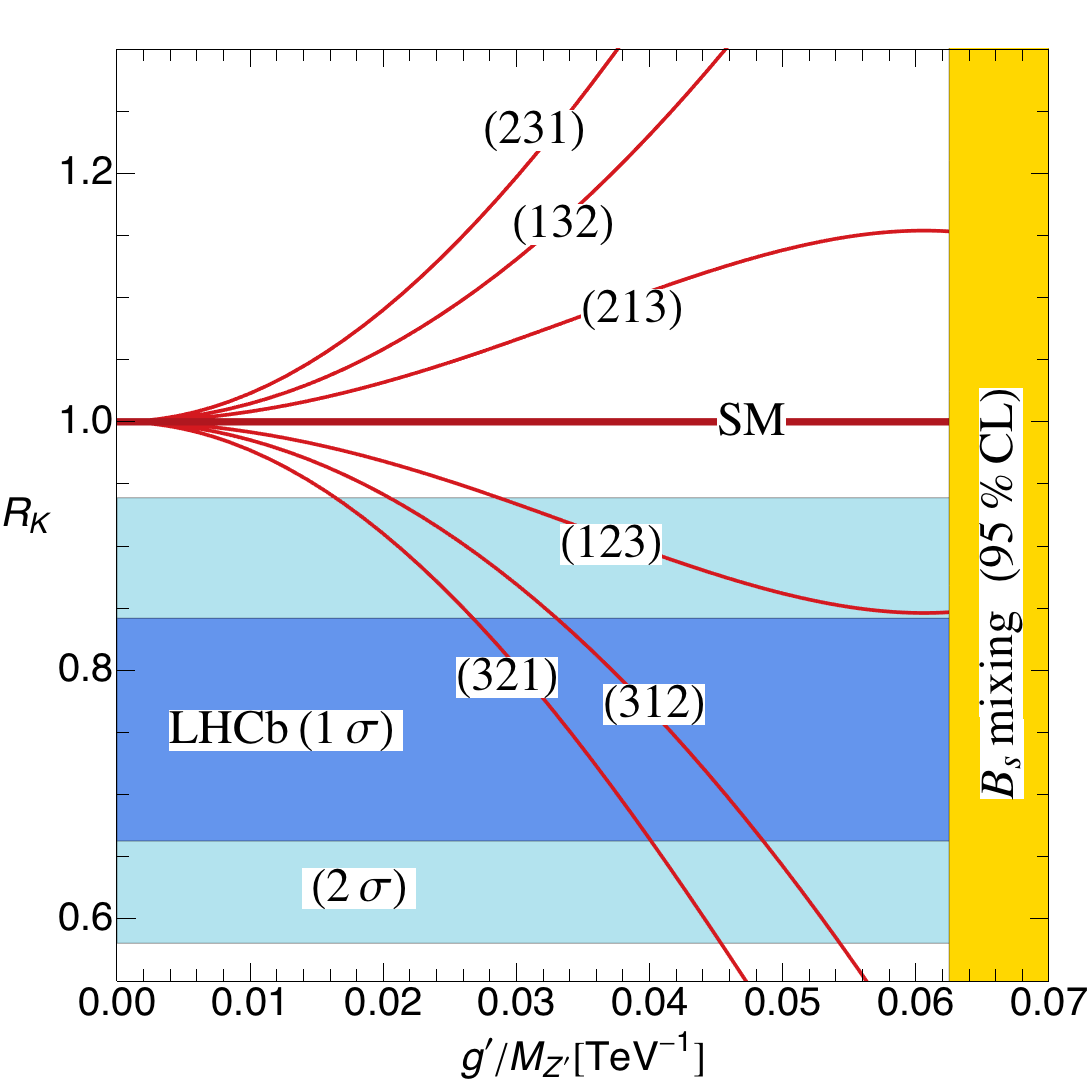}
\end{minipage}
\begin{minipage}{0.49\textwidth}
\centering
\hspace{25pt}
%%%%%%%%%%%%%%%%%%%%%%%%%%%%%%%%%%%%%%%%%%%%%%%%%%%%%%%%%%%%%%%%%%%%%%%%%%
\tabcolsep 0.03in
\begin{tabular}{|c|c|c|c|c|c|} \rowcolor{RGray}
\hline 
Model   & $C_{9}^{\mbox{\scriptsize{NP}}\mu} (1\sigma)$  & $C_{9}^{\mbox{\scriptsize{NP}}\mu}  (2 \sigma)$     \\  \hline 
(1,2,3)   &  -- & $[-2.92, -0.61]$    \\[0,1cm]
(3,1,2) & $[-0.93, -0.43]$  &  $[-1.16, -0.17]$  \\[0.1cm] 
(3,2,1) & $[-1.20, -0.53]$  & $[-1.54, -0.20]$ \\
\hline
\end{tabular}
%%%%%%%%%%%%%%%%%%%%%%%%%%%%%%%%%%%%%%%%%%%%%%%%%%%%%%%%%%%%%%%%%%%%%%%%%%
\end{minipage}
\captionlistentry[table]{}\label{tab:C9_values}
\captionsetup{labelformat=andtable}
\caption{\it \small In figure, model prediction for $R_K$ as a function of $g^\prime/M_{Z^\prime}$. This is shown together with the SM prediction, the experimental measurement by LHCb  at $1\,\sigma$ and $2\,\sigma$ and the bound from $B_s$-mixing. In table, bounds on $C_{9}^{\mbox{\scriptsize{NP}}\mu}$ from $R_K$ for the implementations that are able to accommodate the anomaly.}\label{fig:RKmodels}
\end{figure}
%%%%%%%%%%%%%%%%%%%%%%%%%%%%%%%%%%%%%%%%%%%%%%%%%%%%%%%%%%%%%%%%%%%%%%%%%%

These NP contributions to the effective Hamiltonian can be tested with global fits to the angular distributions of the semileptonic $b\to s\ell^+\ell^-$ transitions. Furthermore, the hadronic ratios
\begin{align}  \label{eq:had_ratios}
R_M \equiv   \frac{\mathrm{Br}(   \bar B \rightarrow \bar M \mu^+ \mu^- )}{\mathrm{Br}(   \bar B
\rightarrow \bar M e^+ e^- )}\stackrel{\rm SM}{=}1+\mathcal O(m_\mu^2/m_b^2)\,,
\end{align}
with $M\in\{K, K^*, X_s, K_0(1430),\ldots\}$~\cite{Hiller:2003js}, provide a precise test on the universality of these transitions. In Figure~\ref{fig:RKmodels} I show the model prediction for $R_K$ from the different implementations of the model together with the recent experimental measurement of the ratio by the LHCb collaboration~\cite{Aaij:2014ora} and the bound from $B_s$-mixing. As we can see, only two of the implementations are able to explain the anomaly at $1\sigma$ and a third one is able to accommodate it at $2\sigma$. For these models, I show in Table~\ref{tab:C9_values} the bounds on $C_9^{\mbox{\scriptsize{NP}} \mu}$ that are extracted from $R_K$. The values obtained for this operator are in good agreement with those favored by the global fits, as was also noticed in other $Z^{\,\prime}$ models~\cite{Altmannshofer:2014rta,Altmannshofer:2015sma,Descotes-Genon:2013wba,Altmannshofer:2013foa,Hurth:2013ssa}. Furthermore, as noted in Ref.~\cite{Hiller:2014ula} the absence of flavor violating NP couplings to right-handed quarks, as it happens in this model, implies a strong condition on the ratios defined in Eq.~\eqref{eq:had_ratios}, $R_K=R_{K^*}=R_{X_s}=\dots$ This provides an important test on the validity of the model and shows the importance of further measurements of these ratios.

Finally, if a $Z^{\,\prime}$ is discovered in the next runs of LHC a useful test on its universality can be found in the ratios
\begin{align}
\mu_{f/f^{\prime}}&\equiv \frac{\sigma(pp\rightarrow Z^{\,\prime}\rightarrow
f\bar{f})}{\sigma(pp\rightarrow Z^{\,\prime}\rightarrow  f^{\prime} \bar  f^{\prime})}\,,
\end{align}
that in our model take the following form
\begin{align}
\mu_{b/t} \simeq  \frac{X_{bL}^2+X_{bR}^2}{X_{tL}^2+X_{tR}^2} \,, \qquad \mu_{\ell/\ell^{\prime}}  \simeq   \frac{X_{\ell L}^2+X_{\ell R}^2}{X_{\ell^{\prime} L}^2+X_{\ell^{\prime} R}^2} \,.
\end{align}
We find $\mu_{b/t}\simeq1$ while the ratios $\mu_{\ell/\ell^{\prime}}$ are highly dependent on the model implementation, opening the possibility to test this model and discriminate among the different implementations.

\section{Summary}\label{Sec:conclusions}
\begin{figure}[t]
\includegraphics[width=4.8cm,height=4.85cm]{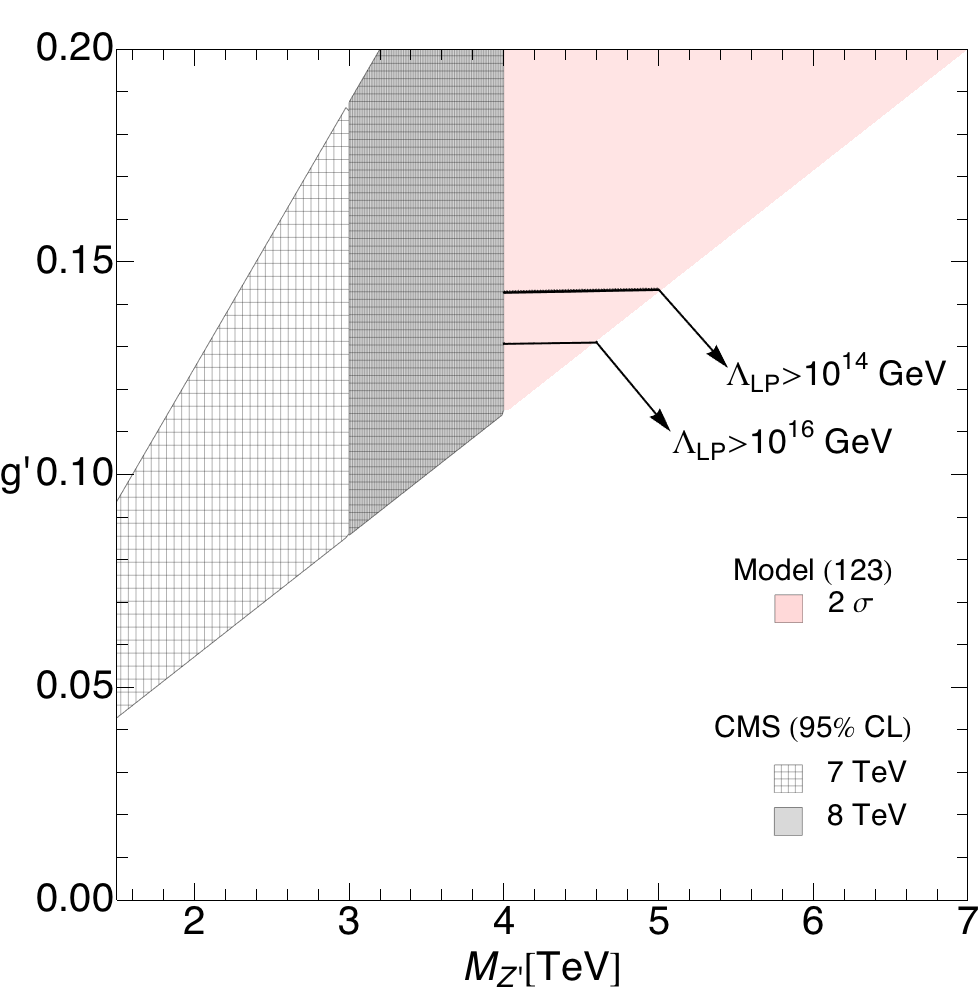}
~
\includegraphics[width=4.8cm,height=4.8cm]{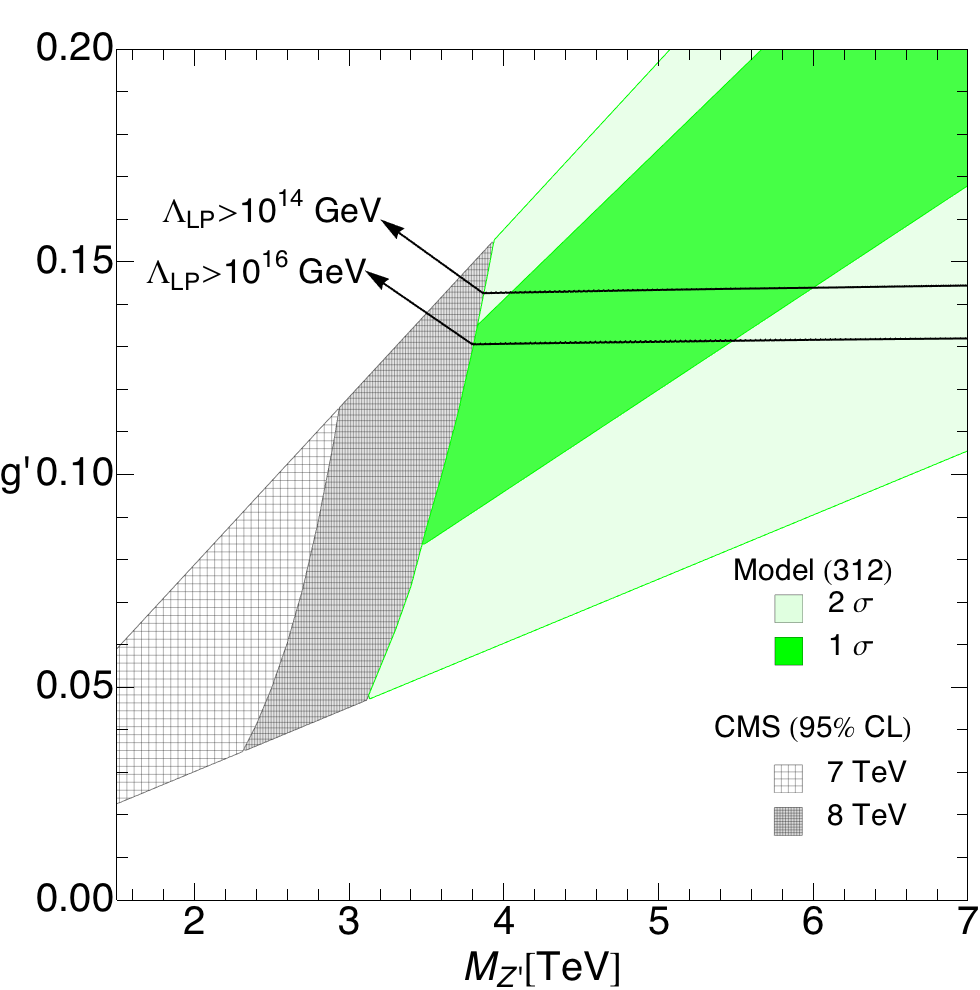}
~
\includegraphics[width=4.8cm,height=4.8cm]{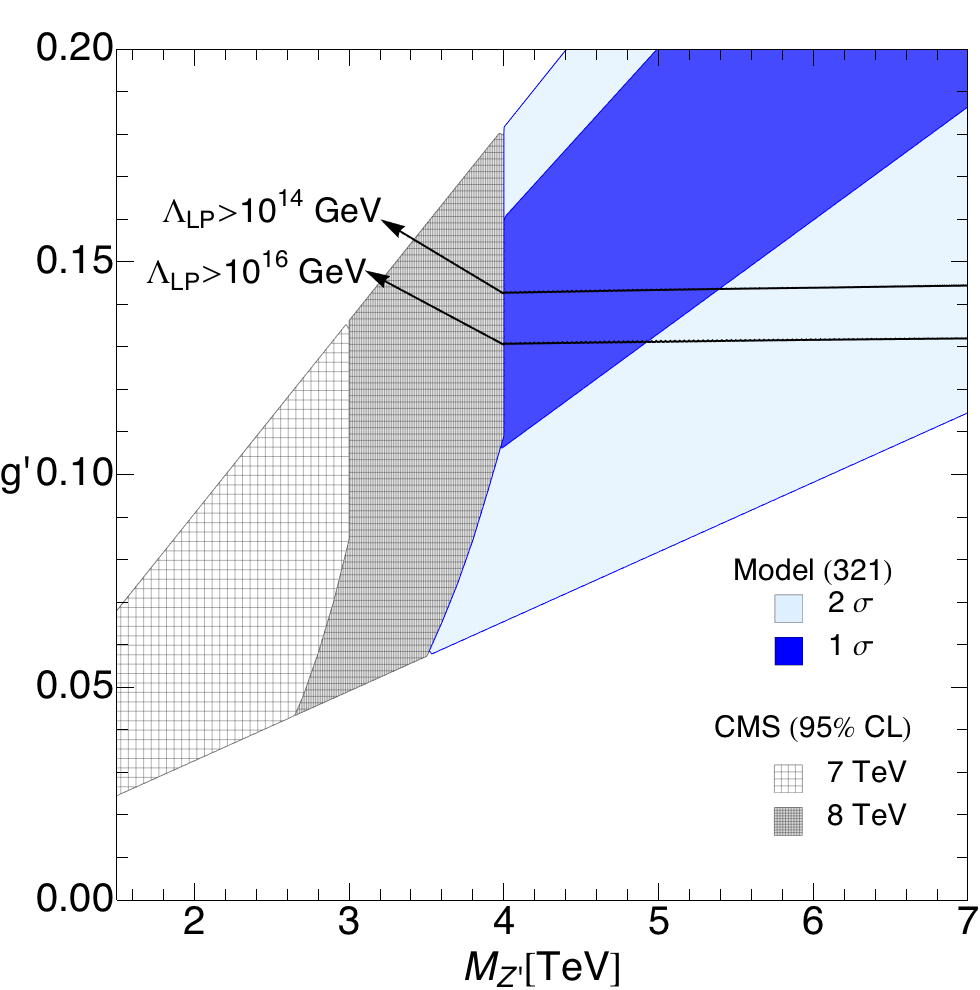}
\caption{\it \small  Regions allowed at $1\sigma$ and $2\sigma$ by the $R_K$ measurement in the
$\{M_{Z^{\prime}},g^{\prime}\}$ plane for the models $(1,2,3)$, $(3,1,2)$ and $(3,2,1)$.   Exclusion
limits from $Z^{\prime}$ searches at the LHC are shown in gray. The black lines indicate bounds from
perturbativity of $g^\prime$.}  \label{fig:constraints}
\end{figure}
In this talk I presented a class of family non-universal $Z^{\,\prime}$ models based on an horizontal gauge symmetry that completely determines the flavor structure of the model, characterized by the presence of tree-level FCNC in the down-quark sector controlled by the CKM matrix, with no flavor violations in the up-quark sector. Anomaly cancellation conditions extend the symmetry to leptons in a precise way, giving rise to flavor-conserving non-universal couplings in the charged-lepton sector and six possible implementations. Moreover, cancellation of anomalies only allows for two free charges which are fixed for phenomenological purposes, leaving only two relevant parameters in the heavy gauge boson sector, the $Z^{\,\prime}$ mass, $M_{Z^\prime}$, and its gauge coupling, $g^\prime$. This renders a highly predictive NP scenario which is able to accommodate the $b\to s\ell^+\ell^-$ anomalies in some of its implementations.

Present data strongly constraints the parameter space of the model: bounds from $B_s$-mixing imply $M_{Z^\prime}/g^\prime\geq 16$~TeV ($95\%$ CL), direct searches at LHC exclude our $Z^{\,\prime}$ for a mass below $3-4$ TeV and perturbativity of the gauge couplings give the upper limit, $g^\prime\lesssim 0.14$. These constraints are shown together with the regions allowed by $R_K$ in Figure~\ref{fig:constraints}. The model also presents smoking-gun signatures that will be tested in the recent future, such as the equality of all the hadronic ratios defined in Eq.~\eqref{eq:had_ratios}, i.e. $R_K=R_{K^*}=R_{X_s}=\dots$ Moreover, if a $Z^{\,\prime}$ is discovered at LHC, measurements of the ratios $\sigma(pp \rightarrow Z^{\,\prime} \rightarrow \ell_i \bar \ell_i )/\sigma(pp\rightarrow Z^{\,\prime} \rightarrow \ell_j\bar \ell_j )$ would be insightful in order to discriminate among the different model variations and from other NP implementations.

\begin{acknowledgments}
I thank the organizers of EPS-HEP 2015 for the opportunity to present these results. I am also grateful to Alejandro Celis, Martin Jung and Hugo Ser\^odio for the collaboration in the topic presented here and for useful comments during the preparation of this manuscript. This work has been supported in part by the Spanish Government, ERDF from the EU Commission and the Spanish  Centro  de  Excelencia Severo  Ochoa  Programme [Grants No. FPA2011-23778, FPA2014-53631-C2-1-P, No.~CSD2007-00042 (Consolider Project CPAN), SEV-2014-0398]. I  also acknowledge VLC-CAMPUS for an ``Atracci\'{o} del Talent"  scholarship.  
\end{acknowledgments}

\end{document}